\documentclass[twocolumn]{aastex701}

\newcommand{\src}{A\,0538$-$66}

\newcommand{\nicer}{{\rm NICER}}

\newcommand{\xmm}{{\rm XMM-Newton}}

\def\msun{{\rm M}_{\odot}}

\begin{document}

\title{Sometimes they come back:   pulsations in \src\ at 69 ms  rediscovered with \nicer\ }

\correspondingauthor{Lorenzo Ducci}
%
%
%

\author[0000-0002-9989-538X]{Lorenzo Ducci} 
\affiliation{Institut f\"ur Astronomie und Astrophysik, Universit\"at T\"ubingen, Sand 1, D-72076 T\"ubingen, Germany}
\email[show]{ducci@astro.uni-tuebingen.de}
 
\author[0000-0003-3259-7801]{Sandro Mereghetti}
\affiliation{INAF, Istituto di Astrofisica Spaziale e Fisica Cosmica di Milano, via Corti 12, I-20133 Milano, Italy}
\email{sandro.mereghetti@inaf.it}

\begin{abstract}
We report on \nicer\ X-ray observations of the Be X-ray binary  \src\ located in the Large Magellanic Cloud. Fast   pulsations (69~ms) in this source  were discovered in 1980 during a bright outburst in which it reached a luminosity of $\sim 8\times 10^{38}$~erg~s$^{-1}$, but were never reobserved since then.
We clearly detected the pulsations at $P=69.3055 \pm 0.0005$~ms with a pulsed fraction of $\sim 20$\%  during a short time interval ($\sim11$ minutes) on 2023 January 9, when   \src\ had a luminosity of $\sim 8\times 10^{36}$~erg~s$^{-1}$ (0.3$–$10~keV). 
The pulsations were not detected in other \nicer\ observations (total exposure $\sim 162.7$~ks), during which \src\ had a similar or lower luminosity.
On   2023 February 8-9  the source exhibited a strong variability, with short flares  reaching $\sim 10^{38}$~erg~s$^{-1}$, but no periodic pulsations were detected. 
Assuming the magnetospheric radius lies within the corotation radius during pulsations, we estimate the neutron star magnetic field is below $\sim 2.7\times 10^{10}$~G. This would make \src\ the high-mass X-ray binary with the weakest known magnetic field. We discuss implications for magnetic field evolution in accreting pulsars and propose that, alternatively, \src\ has a stronger magnetic field and during the \nicer\ detection, a centrifugal barrier may have been active while part of the plasma accumulated at the magnetosphere sporadically leaked through it via an instability mechanism, allowing accretion onto the polar caps.
\end{abstract}



\section{Introduction}
\label{sec:intro}
 
X-ray binaries in which a neutron star (NS) orbits a B-type star with emission lines (Be/XRBs) constitute the majority of the accretion-powered high-mass X-ray binaries (HMXBs) currently known in our Galaxy.  Their X-ray emission is often characterized by a variable and/or transient behavior, caused mainly by the interaction of the NS with the variable circumstellar disk ejected by the donor star in its equatorial plane (see, e.g., \citealt{2011Ap&SS.332....1R}, for a review of Be/XRBs).

Among Be/XRBs,  \src, located in the Large Magellanic Cloud, stands out for some quite unique properties. The NS in this system has a spin period of only 69~ms \citep{1982Natur.297..568S} and is in a very eccentric orbit (e$\sim$0.72) with a period of $\sim16.6$ days around its Be companion (\citealt{2017MNRAS.464.4133R}, \citealt{1979ApJ...230L..11J}).  Modulation at the orbital period is visible at optical and X-ray energies \citep{1980Natur.288..141S}, with the brightest and fastest optical flares observed in HMXBs, likely powered by a reprocessing mechanism (\citealt{2019A&A...624A...9D} and references therein).

The fast X-ray pulsations in \src\  were discovered during a very bright outburst observed with the {\it Einstein Observatory} in December 1980, when the source reached the super-Eddington luminosity of $\sim8\times 10^{38}$~erg~s$^{-1}$. The pulsed fraction was $\sim$26\% and a rapid spin-down of $5\times10^{-10}$~s~s$^{-1}$, consistent with the effect of  orbital motion in an eccentric orbit, was measured \citep{1982Natur.297..568S}.  Since then, despite several attempts,  the pulsations were never detected again \citep{1993A&A...274..304M,1997ApJ...476..833C,2002ApJ...580..389C,2019ApJ...881L..17D},
possibly because all the subsequent observations caught \src\ at lower luminosity levels (see, e.g., \citealt{2025AN....34640098R}).

The Neutron Star Interior Composition Explorer (\nicer ) repeatedly observed \src\ from 2022-11-17 and 2025-04-17. We used these data to perform a sensitive search for periodicities exploiting their high counts statistics.
We present the data analysis and results in Section~\ref{sec:observation} and we discuss their main implications in Section~\ref{sec:discussion}.
All the results are given for a source distance of 50~kpc.

\section{Observations, data analysis and results} 
\label{sec:observation}

We selected all the observations of \src\ present in the public archive of \nicer\ data. They consist of 74 observations 
amounting to a total on-source exposure time of 162.7~ks  (see Table \ref{table log}).
We followed the standard procedures for data reduction and the filtering criteria provided by the \nicer\ team on the HEASARC website\footnote{\url{https://heasarc.gsfc.nasa.gov/docs/nicer/analysis_threads/}}. 
Given the faint and bursting nature of \src, we applied more conservative filtering following recommendations in \nicer\ threads, excluding times with cutoff rigidity values below 1.5~GeV/$c$ and limiting ``overshoot'' rates to below 5~ct/s to reduce contamination from particle-induced flares\footnote{For more information, see: \url{https://heasarc.gsfc.nasa.gov/docs/nicer/analysis_threads/flares/}}.
We used the latest HEASOFT software (version 6.35.1), calibration database (CALDB release date 2024-02-27), and the geomagnetic data from \nicer\ SOC (installed: 2025-06-09)\footnote{\url{https://heasarc.gsfc.nasa.gov/docs/nicer/analysis_threads/geomag/}}.
The arrival times  were converted to the solar system barycenter using the tool {\tt barycorr} and the JPL planetary ephemeris DE-440.

As a first step, we carried out a search for periods in the range 0.04$-$1~s in each individual observation using the whole energy range (0.2$-$15~keV). The search was done with the Rayleigh ($Z^2_n$) test using one harmonic ($n=1$; see e.g., \citealt{Buccheri83}), sampling only the statistically independent periods in each observation.

\begin{figure}
\begin{center}
\includegraphics[width=0.9\columnwidth]{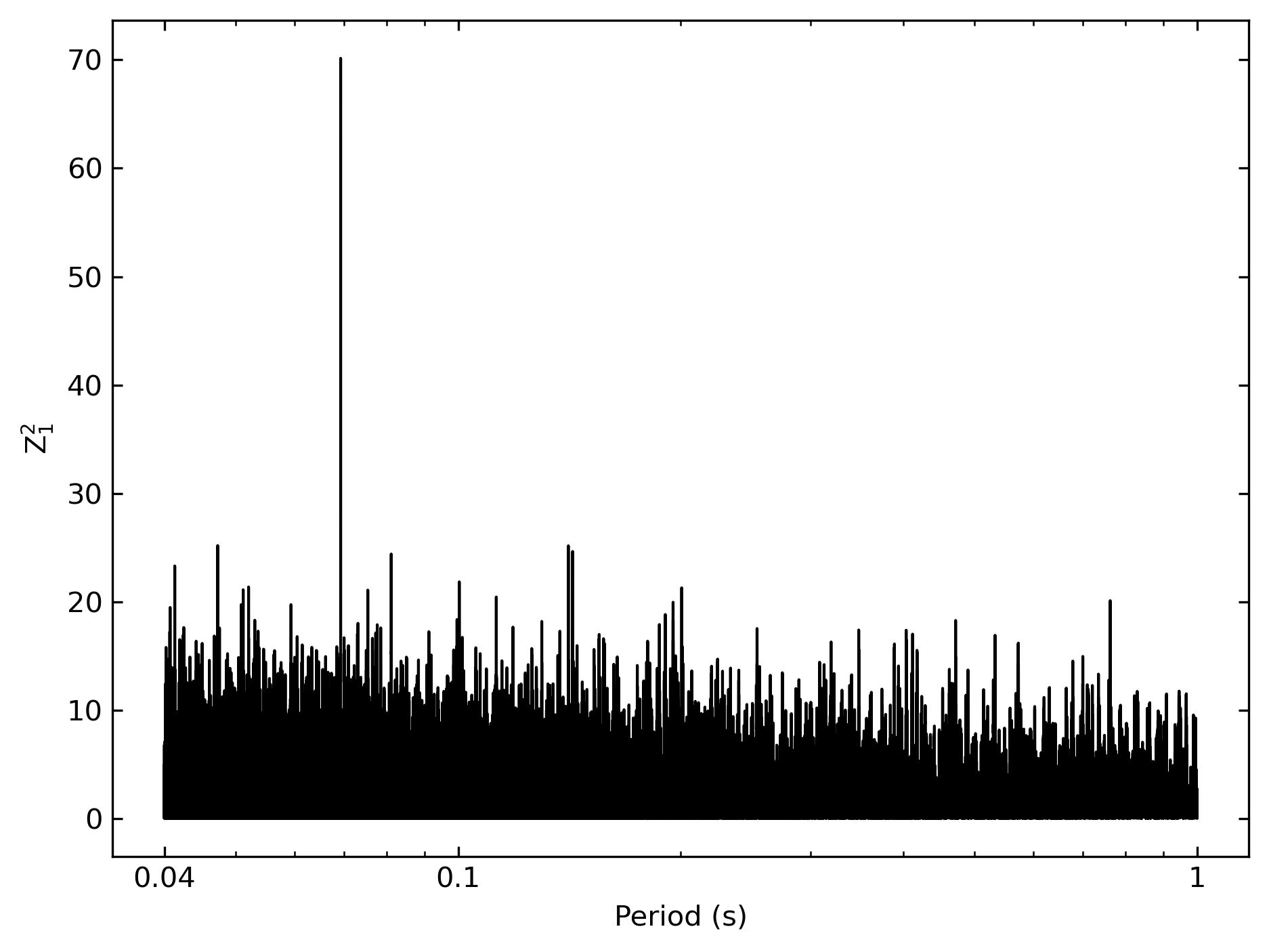}
\caption{Rayleigh test computed for the ObsID 5203560125 of \src, without applying energy selection. 
\label{fig:Z2}}    
\end{center}
\end{figure}

The only significant signal was found in ObsID 5203560125, with  the test statistics Z$^2$ = 70.14  for the period $P \approx 0.0693$~s (see Fig. \ref{fig:Z2}). 
The single trial probability that this is a chance fluctuation is $5.8\times 10^{-16}$, and, taking into account the total number of examined periods  
(110,881,248\footnote{This is a conservative estimate because many of the trial periods are common to more than one observation.}, see Table \ref{table log}), we obtain a probability of $6.5\times 10^{-8}$. 
Excluding three peculiar observations\footnote{In ObsID 6203560125 strong $Z^2$ peaks are present at exact integer frequencies (2, 3, 4, 6, 7, ..., 12~Hz), which strongly suggests an instrumental origin, while 
in observations  5203560138 and 5203560139 there are prominent $Z^2$ peaks at low frequencies, due to the strong flaring activity from the source discussed below.}, the next highest peak was found at $P\approx0.0587$~s in ObsID 5203560110, with a value of the test statistics ($Z^2$=37.99) fully compatible with a chance  fluctuation.

\begin{figure*}
\begin{center}
\includegraphics[width=2.0\columnwidth]{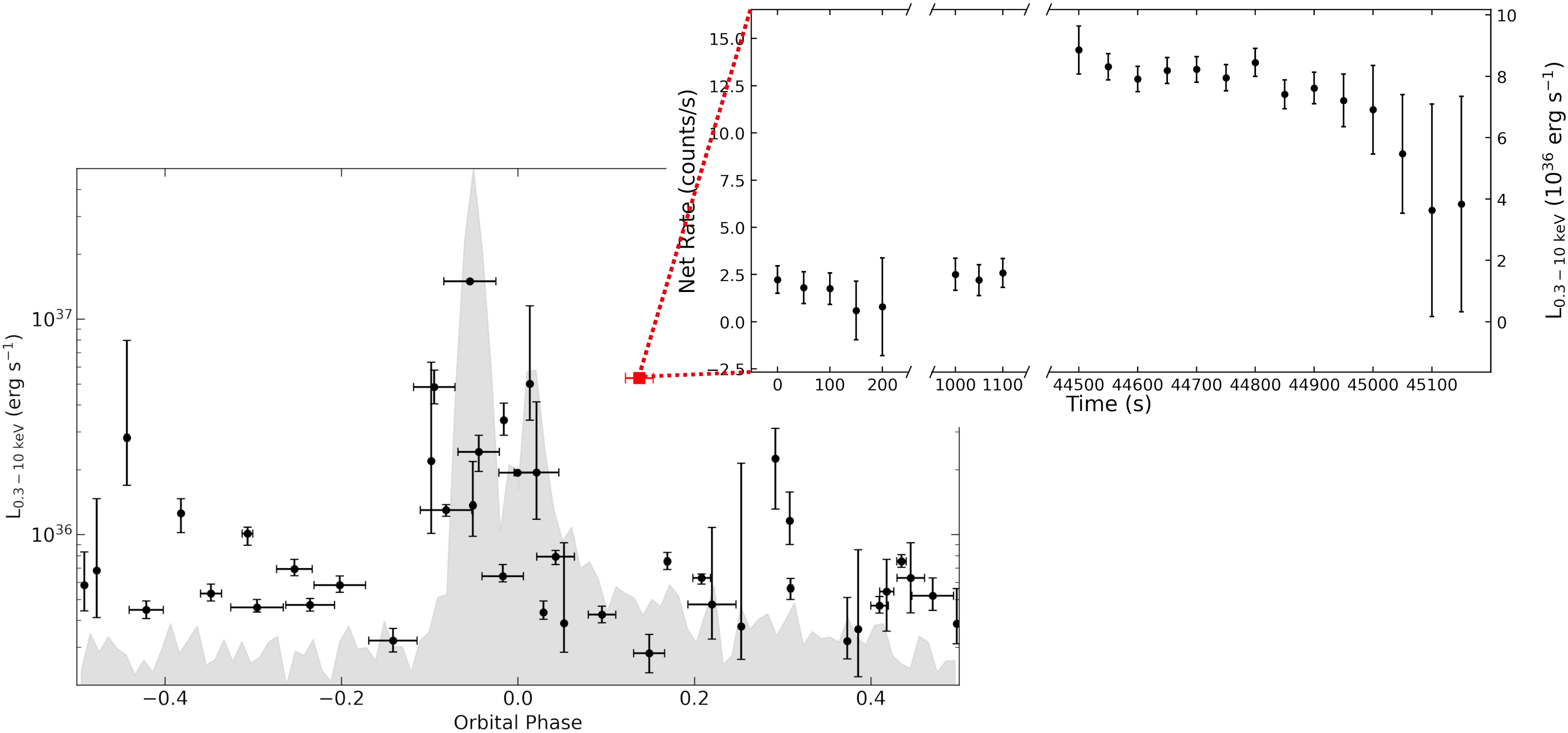}
\caption{\emph{Left:} \nicer\ X-ray light curve of \src\ folded with the orbital period (orbital phase zero is the periastron passage; see \citealt{2022A&A...661A..22D}, \citealt{2017MNRAS.464.4133R}). 
The phase-folded data cover 53 orbital cycles, producing scatter in $L_{\rm x}$ due to cycle-to-cycle variability.
The gray band shows the (renormalized) optical orbital profile from \citet{2022A&A...661A..22D}. This profile is known to be in-phase with the X-ray modulation and is shown for reference.
X-ray luminosities (0.3$-$10~keV) are obtained from spectral analysis and assuming $d=50$~kpc. X-ray pulsations were detected in the observation marked by the red square. \emph{Right:} Zoom of the observation with the periodic signal: luminosity increases in the third time interval where pulsations are detected.
\label{fig:lcr}}    
\end{center}
\end{figure*}

\begin{figure}
\begin{center}
\includegraphics[width=0.9\columnwidth]{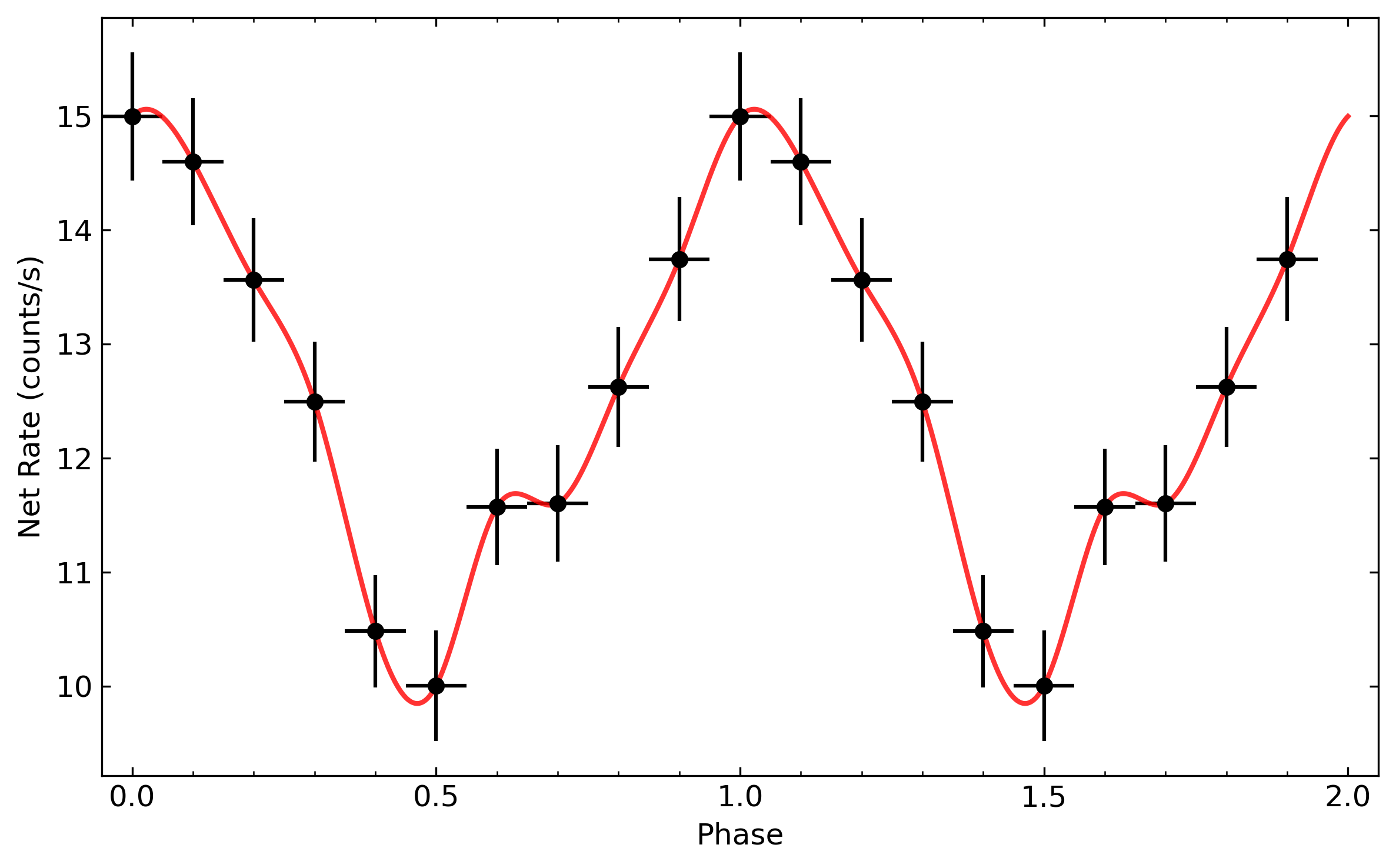}
\caption{0.4$-$13~keV pulse profile of \src\ in the last time interval of ObsID~5203560125 ($\phi=0$ at $T=59953.0$~MJD).
Black points with error bars show binned measurements; red curve traces a smoothed representation.
\label{fig:pprofile}}    
\end{center}
\end{figure}

A closer examination of ObsID~5203560125 showed that  the periodic signal is present only in the last time interval (duration of 662.5~s), when the source was a factor $\sim$5 brighter (see inset of Fig. \ref{fig:lcr}).
The maximum significance of the periodic signal ($Z^2=94.85$) was obtained by selecting only the counts of this time interval in the 0.4$-$13~keV energy range\footnote{Given the small  effective area above 12 keV, this enhanced significance might be due to a statistical fluctuation;  reducing the upper energy boundary  we still obtain highly significant detections:  $Z^2=92.55$ (0.4-12 keV), $Z^2=93.05$ (0.4-11 keV).}.
To obtain a more precise period measurement we performed a phase fitting of the folded light curves obtained by subdividing the data into three time segments. The resulting period is  $P=69.3055 \pm 0.0005$~ms and the corresponding background-subtracted pulse profile in the 0.4$-$13 keV band is shown in Fig. \ref{fig:pprofile}.
The   pulsed fraction, defined as $p_{\rm f}=(F_{\rm max} - F_{\rm min})/(F_{\rm max} + F_{\rm min})$ is 20$\pm$3\%.
Following \citet{2002ApJ...575L..21M, 1994MNRAS.268..709B}, we derived a 3$\sigma$ upper limit (u.l.) on the pulsed fraction in the first part of the observation
of $p_{\rm f}\lesssim$46\%.

Based on these findings, we made a new search restricted to the period range 0.067$-$0.071~s and the energy range 0.4$-$13~keV  in  all the individual observations, as well as splitting each observation in sub-intervals with elapsed time not exceeding 3~ks.
No significant signals were found, besides that of ObsID 5203560125.
The search was also repeated in the 0.4$-$12~keV and  0.4$-$10~keV energy ranges, which could in principle reduce the background contribution, but again   negative results were obtained.
To gain more exposure for pulsation searches, we relaxed the event filtering criteria to default values. In ObsID 5203560125, this increased exposure during the faint, early phase of the observation, but no pulsations were detected there, nor in any other observation.
Due to the different durations of the observations and variable flux of the source, the upper limits on the pulsed fraction are in the range  $\sim5-100$\%. It is important to note that at least   other seven observations had a sensitivity sufficient to reveal a pulsed signal similar to that seen in  ObsID 5203560125.

\begin{figure*}
\begin{center}
\includegraphics[width=1.9\columnwidth]{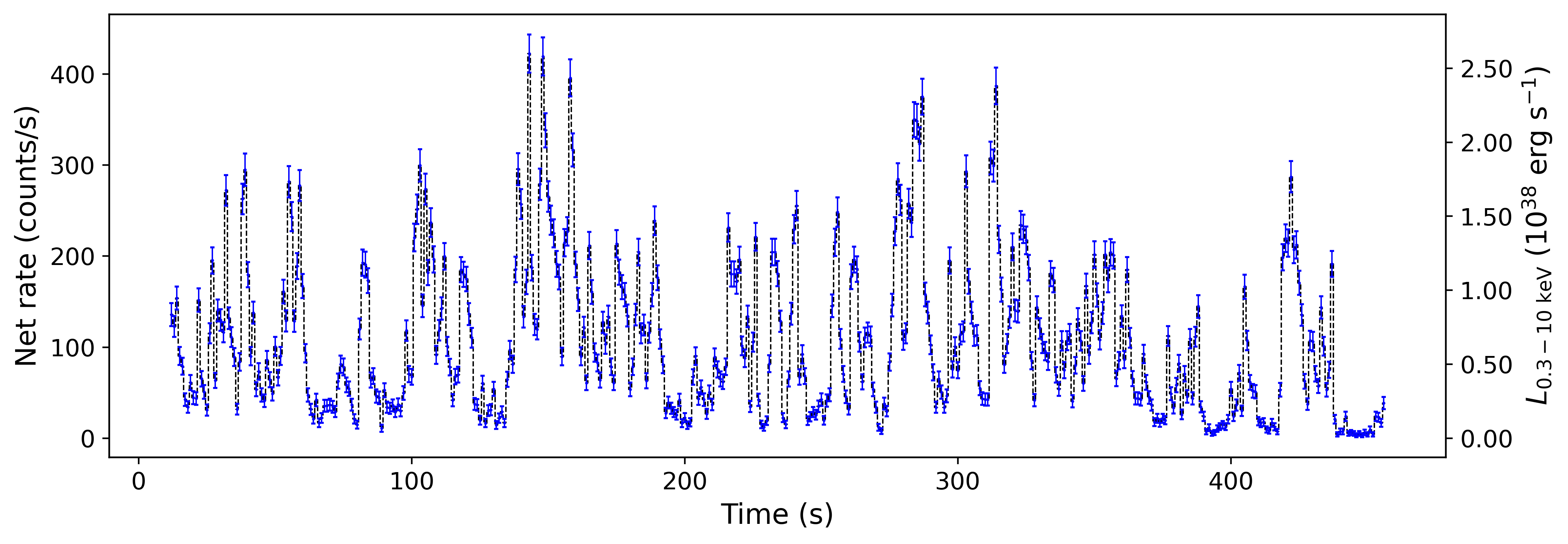}
\caption{
Light curve of \src\ (0.3$-$10~keV) during a part of ObsID 5203560138 showing a strong flaring activity. The time bin is 1 s. 
Blue markers with error bars show the net count rate (left axis) and their $1\sigma$ uncertainties, connected by a black dashed step function to trace the temporal evolution. The right axis shows the corresponding luminosity.
The background contribution was estimated using BACKV and BACKE values provided in output lightcurves by the task {\tt nicerl3-lc} using the option {\tt bkgmodeltype=scorpeon}.
\label{fig:lcr 520...138}}    
\end{center}
\end{figure*}

The observations 5203560138 and 5203560139  are particularly noteworthy. They occurred at orbital phase $\phi \sim 0$. In ObsID 5203560138, \src\ showed its highest average  luminosity (see Fig. \ref{fig:lcr}). A closer inspection of the 0.3$–$10~keV light curves with 1-second time bins in these two observations revealed strong fast  variability resembling the flaring state seen in two   \xmm\ observations carried out on 2018 May 15 and 31   at the  phase $\phi \sim 0$ of two consecutive   orbits \citep{2019ApJ...881L..17D}. Figure \ref{fig:lcr 520...138} displays a representative flaring episode from the  ObsID 5203560138. Six such episodes occurred during this observation, and three in ObsID 5203560139. The flares in Fig. \ref{fig:lcr 520...138} are very short, span nearly two orders of magnitude in flux, and reach a peak luminosity of $\sim 2.5\times 10^{38}$~erg~s$^{-1}$. In these observations, we obtained
a $3\sigma$ u.l. on $p_{\rm f}$ of $\sim 5$\%. However, we caution that this negative result might be caused by the extreme flaring behavior of the source.
We defer a detailed analysis of these observations to a forthcoming publication.

For each \nicer\ observation, we performed spectral extraction and background modeling using the SCORPEON method\footnote{\url{https://heasarc.gsfc.nasa.gov/docs/nicer/analysis_threads/scorpeon-overview/}}. Spectra with sufficient source signal were then analyzed with XSPEC (version 12.15.0). Absorption was modeled with the {\tt tbabs} component using \citet{2000ApJ...542..914W} abundances and the photoionization cross sections from \citet{1996ApJ...465..487V}. Following the recommendations of the \nicer\ team, we adopted the Poisson {\tt pgstat} statistic for the fit, excluded energy channels below 0.22~keV and above 15~keV, and incorporated systematic errors assigned to each spectral channel through the automated {\tt niphsyserr} task in {\tt  nicerl3-spect}. 
The SCORPEON model parameter ranges and settings are custom tailored to observational conditions; therefore, as advised by the \nicer\ team, we did not modify them. 

\begin{figure*}
  \centering
  \begin{minipage}[c]{0.95\columnwidth}
    \includegraphics[width=\textwidth]{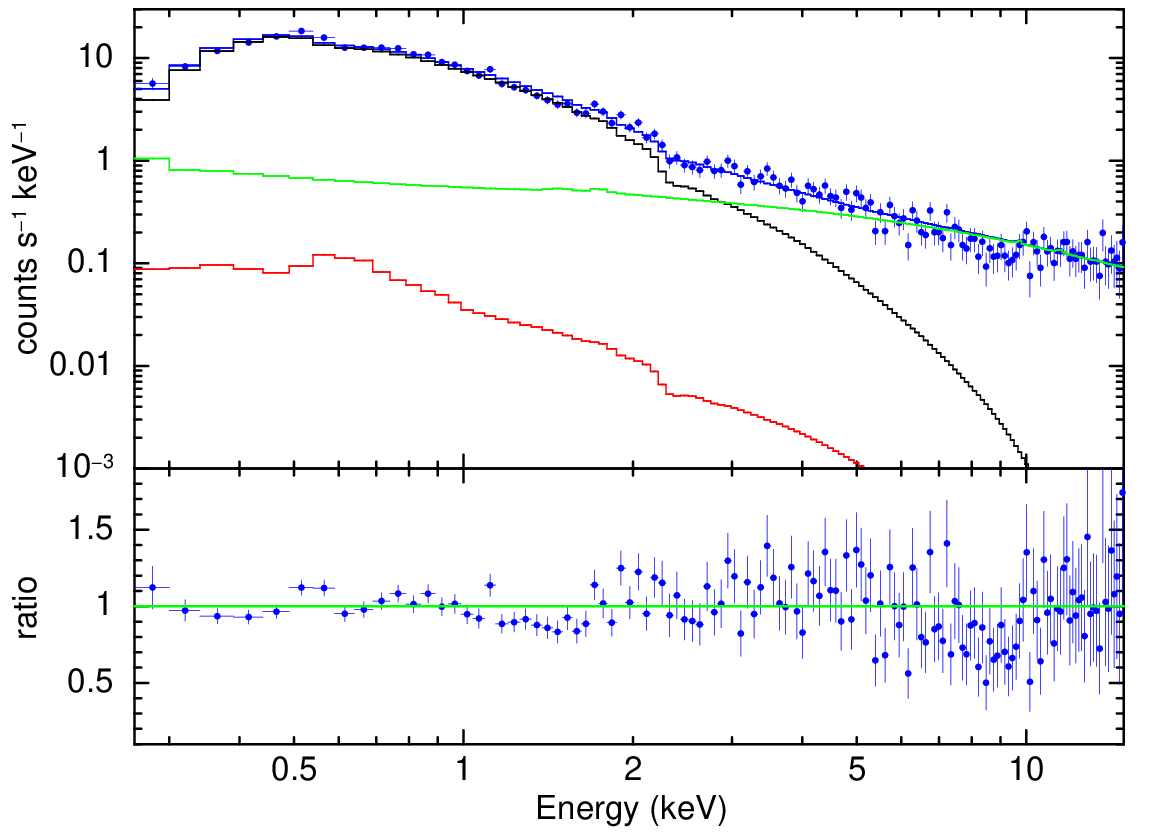}
  \end{minipage}
  \begin{minipage}[c]{0.95\columnwidth}
    \includegraphics[width=\textwidth]{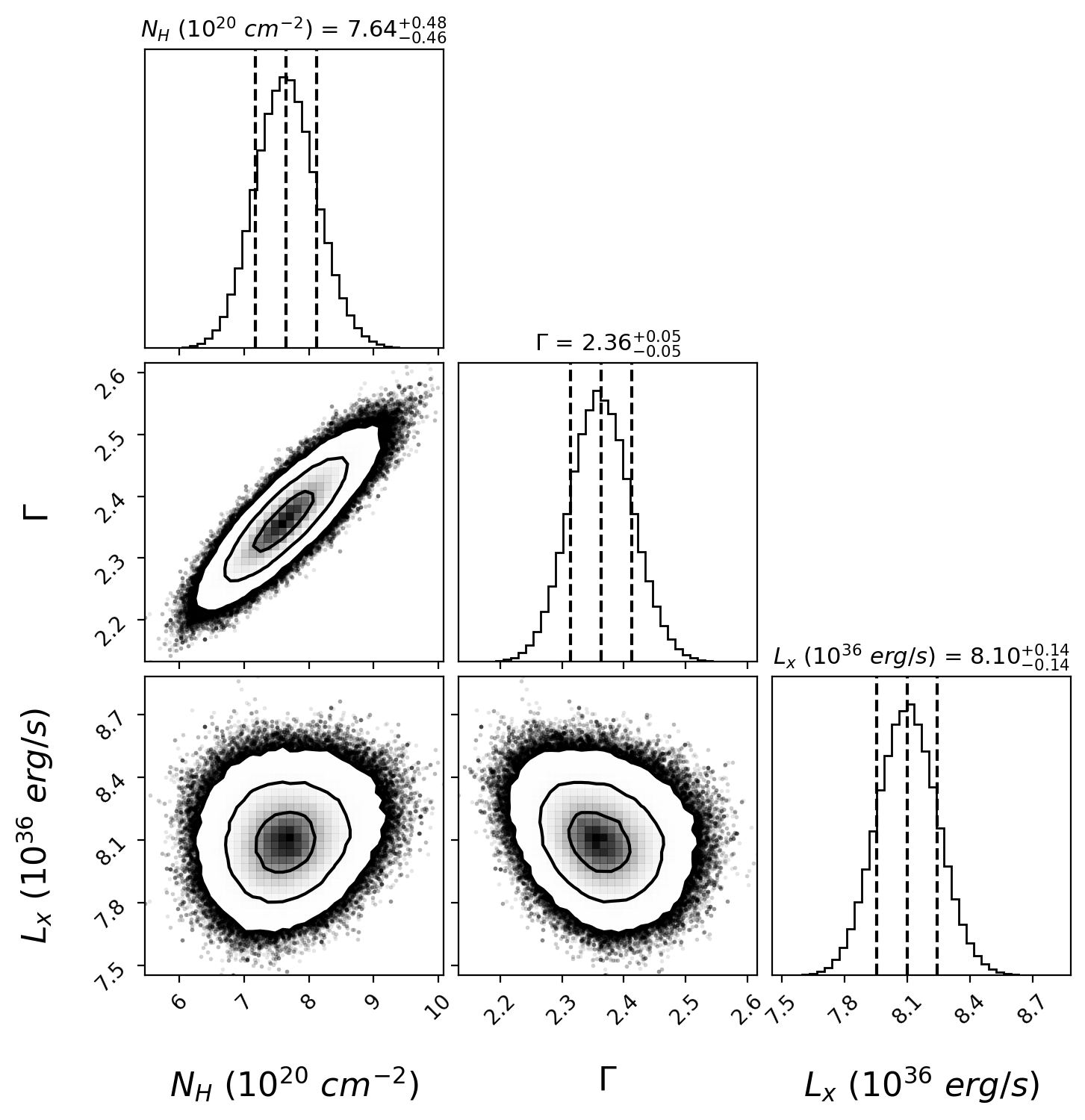}
  \end{minipage}
  \caption{\emph{Left panel}: best fit spectrum (top) and residuals (bottom) of the last time interval of ObsID~5203560125, where pulsations are detected. Data (blue points) are shown with the total best-fit model (blue line), source model (black line), non-X-ray background (green), and diffuse X-ray sky background (red). \emph{Right panel}: Corner plot of spectral parameters from MCMC simulations. Contours indicate 1, 2, and 3$\sigma$ confidence levels in 2D distributions. $L_{\rm x}$ is the unabsorbed 0.3$-$10~keV luminosity.
  \label{fig:spectrum}}
\end{figure*}

For all the spectra, we obtained good fits with an absorbed power law model and calculated the 0.3$-$10~keV unabsorbed fluxes using the {\tt cflux} convolution model. The resulting X-ray luminosities are shown in Figure \ref{fig:lcr}.
For the  spectrum of the last time interval of ObsID 5203560125, where the pulsating signal was detected, we obtained the best fit (pgstat/d.o.f. = 189.82/130; null hypothesis probability $1.19\times 10^{-3}$) 
with column density  $(7.6\pm 0.5)\times 10^{20}$~cm$^{-2}$,  photon index  $\Gamma=2.36 \pm 0.05$, and
unabsorbed flux $(2.71{+0.04\atop -0.05})\times 10^{-11}$~erg~cm$^{-2}$~s$^{-1}$.  
The best-fit spectrum is shown in Fig. \ref{fig:spectrum}, left panel. 
We performed Markov Chain Monte Carlo (MCMC) sampling to examine global parameter relationships and degeneracies (Fig. \ref{fig:spectrum}, right panel) and to determine  the uncertainties for the best-fit spectral parameters reported above. Using the XSPEC Goodman-Weare algorithm, we ran 20 walkers for $2\times10^6$ iterations following a $10^6$ iteration burn-in phase\footnote{See the XSPEC manual for details: \url{https://heasarc.gsfc.nasa.gov/docs/xanadu/xspec/manual/manual.html}}.

The spectrum shows a minor excess around $\sim 0.5-0.6$~keV. To investigate its potential origin, we tested relaxing the normalization parameters on emission lines frozen to zero in the standard SCORPEON background script. We found that freeing the normalization parameter of the O~VII line resulted in only a small improvement to the fit quality. The best-fit statistic with O~VII normalization free is pgstat/d.o.f. $=182.26/129$ (null hypothesis probability: $3.1\times 10^{-3}$). Given this negligible improvement,  and considering that the main goal of our spectral analysis is to estimate the source luminosity, we kept the standard configuration (with O~VII normalization frozen to zero) for our final analysis.

 \section{Discussion} 
 \label{sec:discussion}

Our   significant detection of pulsations at 69~ms in the X-ray emission of \src\ is the first one after the discovery of this periodicity, more than four decades ago \citep{1982Natur.297..568S}.
The pulse profile observed with \nicer\ on  2023 January 9, characterized by a single broad pulse with a 
potential secondary peak or ``shoulder'' on the rising part, is remarkably similar to that observed in 1980.
At face value,  the current spin period, longer than that seen in the  $Einstein ~Observatory$ data ($P=69.21166 \pm 0.00017 $~ms on  JD = 2,444,590.4792), would imply an average spin-down rate of  $\dot{P} = (7.07 \pm 0.04)\times10^{-14}$~s~s$^{-1}$. 
However, the   difference between the two periods ($\Delta P/P=1.35\times10^{-3}$) is fully compatible with that caused by the orbital motion of the NS.
In fact, given the large eccentricity $e=0.72\pm0.14$, and plausible values for the other system parameters (inclination $i<75^{\circ}$, companion mass  $10 \msun \lesssim M_c \lesssim 15 \msun$) derived by \citet{2017MNRAS.464.4133R}, a Doppler shift up to $\Delta P/P = V_{\rm rad}/c \sim (1-2)\times10^{-3}$ could affect the 1980 period measurement.
Unfortunately, the current knowledge of the system's orbital parameters is not sufficiently accurate to correct the arrival times for the orbital motion and thus derive the intrinsic long-term  evolution of the NS spin period. 

The spin period of \src\  is the shortest one\footnote{A possible period of 34~ms has been reported in SAX~J0635+0333 \citep{2000ApJ...528L..25C}, but it has not been confirmed and it is not clear yet if this source is accretion- or rotation-powered \citep{2009A&A...504..181M}. } among accretion-powered NSs in HMXBs.
Accretion down to the surface of a NS can occur only if the matter inflow rate is sufficiently strong to overcome the centrifugal barrier produced by the rotating magnetosphere \citep{1975A&A....39..185I}. In other words, the magnetospheric radius ($R_{\rm m}$, defined as the distance from the NS where the magnetic pressure and the ram pressure of the gravitationally captured matter are in equilibrium)  must be smaller than  the corotation radius ($R_{\rm co}$, where the Keplerian orbital period matches the NS spin).  
The presence of pulsations at 69~ms, when the source had a luminosity approaching 10$^{39}$~erg~s$^{-1}$, was used to  derive an upper limit of the order of 10$^{11}$~G  on the magnetic field strength of the NS in \src\ \citep{1982Natur.297..568S,1995A&A...297..385C}. This limit is rather low, but still compatible with what is expected in a young NS.  Applying the same argument to the \nicer\ observation, in which pulsations were seen at a luminosity of only $8\times10^{36}$~erg~s$^{-1}$, yields a much lower upper limit $B_{\rm max}$ on the magnetic field. 
In fact, using standard definitions of $R_{\rm m}$ and $R_{\rm co}$    \citep{1981MNRAS.196..209D}
and the relation for the X-ray luminosity $L_{\rm x} \approx G M_{\rm NS}\dot{M}_{\rm acc}/R_{\rm NS}$, we obtain:
\begin{equation} \label{eq:Blim}
B_{\rm max} \approx 3\times 10^{10} L_{37}^{1/2} R_6^{-5/2} M_{1.4}^{1/3} P_{69}^{7/6} \mbox{\ G,}
\end{equation}

\noindent
where $\dot{M}_{\rm acc}$ is the mass accretion rate,  $L_{37}$ is the luminosity in units of $10^{37}$~erg~s$^{-1}$, $R_6$ is the NS radius in units of $10^6$~cm,
$M_{1.4}$ is the NS mass in units of 1.4~M$_\odot$, $P_{69}$ is the spin period in units of $69$~ms.
To our knowledge, this would be the lowest dipolar magnetic field strength in a HMXB, more than one order of magnitude below the second lowest one of
$\lesssim 5\times 10^{11}$~G in MCSNR~J0513-6724 \citep{2019MNRAS.490.5494M}. 

Such a low field contrasts with typical birth magnetic fields of NSs, 
which follow a lognormal distribution with $\log_{10} [B/G] = 12.44\pm0.44$ \citep{2022MNRAS.514.4606I}. 
Alternatively,  the low field could be caused by suppression via accretion of an originally stronger field (e.g. \citealt{1974SvA....18..217B}, \citealt{2024Galax..12....7A}), as it has been proposed to explain the weak fields  ($\sim10^{10}–10^{11}$~G) of Central Compact Objects -- isolated NSs  with spin periods of $0.1-0.4$~s found in young supernova remnants (see, e.g., \citealt{2017JPhCS.932a2006D}, \citealt{2023Univ....9..273P}). However, also this scenario is unlikely to apply for \src\ since, after the accretion episode, the strong internal field is expected to reemerge on a short time scale of $10^3-10^4$~yr, \citep{2007MNRAS.376..609P, 2011MNRAS.414.2567H} while both evolutionary scenarios   \citep{1998A&A...340...85R} and the association  
with the open cluster NGC~2034 \citep{1981A&A...102L...1P}, indicate that \src\ has a most likely age of several million years.

An alternative possibility is that the NS magnetic field exceeds the upper limit estimated with Eq.~\ref{eq:Blim}. In this scenario, \src\ would have been observed pulsating by \nicer\ during a centrifugal inhibition regime, an unexpected occurrence given that such conditions typically suppress accretion.
The detection of pulsations could be explained by a rare event in which matter accumulates near the magnetospheric boundary
and suddenly leaks through it, due to some instability mechanism, reaching the NS surface and producing pulsed emission.
Similar episodic accretion events have been proposed to explain X-ray variability in several accreting NSs \citep[e.g.,][]{2008ApJ...683.1031B, 2010MNRAS.406.1208D, 2019ApJ...881L..17D} and the detection of pulsations in some high-mass X-ray binaries during low-luminosity states \citep[e.g.,][]{2008A&A...480L..21P, 2014A&A...561A..96D, 2017MNRAS.470..126T}. These mechanisms require instabilities, such as gravitational interchange instabilities or rapid transitions between direct accretion and a centrifugally inhibited accretion regime, allowing a small fraction of plasma to reach the NS surface. Specifically, some models imply the presence of a dead accretion disc acting as a matter reservoir around the NS during the centrifugal inhibition regime. If matter continues to reach the outer disc radius during low-luminosity states, intermittent plasma ingress through the centrifugal barrier at the inner disc radius might occur, producing enhanced X-ray luminosity and pulsations. 
If accretion down to the NS surface occurs through instability events, it is also possible that 
the plasma follows different magnetic field lines which only rarely lead to a favorable beaming of the emission relative to the observer's line of sight, 
thus explaining the absence of pulsations in most observations.
Other models invoke stable accretion at low luminosity levels from cold disc ($\sim 6500$~K) around the NS, which can also allow the matter to reach the NS poles \citep[see discussion in][]{2017MNRAS.470..126T}.

The interpretation that accretion instabilities plays a significant role in the behavior of \src\ 
is also supported by the rapid and strong variability displayed by \src\ on 2023  February 8-9 (Fig.~\ref{fig:lcr 520...138}) and in two \xmm\ observations close to periastron \citep{2019ApJ...881L..17D}. As discussed by these authors, such a behavior is unique among Be/XRBs and can be explained if the source is surrounded by a nearly spherically symmetric  inflow leading to the formation of an atmosphere above the magnetosphere and to rapid variability related to transitions between the regimes of supersonic propeller and direct accretion.

\section{Conclusions}

Thanks to the large collecting area and high time resolution of \nicer , we obtained  the first confirmation of the 69~ms pulsations in \src , that were observed only once in 1980 when the source had a super-Eddington luminosity of $\sim 8\times 10^{38}$~erg~s$^{-1}$. Remarkably, the pulsations were seen in one of the \nicer\ observations showing the highest luminosity, but still two orders of magnitude below that of the 1980 outburst. 

If the NS in \src\ was in the regime of direct accretion ($R_{\rm m} < R_{\rm co}$), the presence of pulsations at this luminosity level implies an upper limit of $\sim 2.7\times 10^{10}$~G on its dipole magnetic field, about ten times below the previous upper limit and surprisingly small for a young NS in a HMXB. 

We propose that an alternative scenario, not requiring such a low magnetic field,  and also supported by the very sporadic nature of the pulsations and by the strong and rapid variability  unique to \src , is that the pulsations occurred during a rare episode in which matter leaked through the centrifugal barrier. 

In any case, the rediscovery of fast pulsations in \src\ is promising for future observations which, coupled to a better determination of the orbital parameters, will permit to study the torques acting on this fast rotating NS and to better understand the peculiar behavior of \src .

\begin{acknowledgments}
LD acknowledges funding from German Research Foundation (DFG), Projektnummer 549824807.
SM acknowledges financial support through the INAF grants ``Magnetars'' and ``Toward Neutron Stars Unification''.
\end{acknowledgments}

%


\facilities{ NICER}

\software{{\tt stingray} \citep{Bachetti22, Huppenkothen19a, Huppenkothen19b},
          {\tt astropy} \citep{astropy:2013, astropy:2018, astropy:2022},
          {\tt XSPEC}   \citep{Arnaud96}.
 }





\bibliography{A0538}{}
\bibliographystyle{aasjournalv7}



\appendix
\renewcommand{\thetable}{A.\arabic{table}}  
\setcounter{table}{0}  


\begin{table*}[h!]
\centering
\caption{Observation log}
\label{table log}
\small 
\begin{minipage}[t]{0.49\textwidth}
\centering
\resizebox{\textwidth}{!}{
\begin{tabular}{lrccrr}
\hline
\hline
ObsID & $T_{\rm exp}$ & \multicolumn{1}{c}{Start time} & \multicolumn{1}{c}{Stop time} & $N_{\rm cts}$ & $N_{\rm p}$ \\ 
      & (s)           & (UTC)                         & (UTC)                        &               &             \\
\hline
5203560101  &  590  &  2022-11-17T22:35:09  &  2022-11-17T22:44:59  &  2646  &  28316  \\
5203560102  &  1030  &  2022-11-18T03:18:44  &  2022-11-18T23:30:26  &  4070  &  3489665  \\
5203560103  &  632  &  2022-11-19T00:57:00  &  2022-11-19T04:08:48  &  2473  &  552368  \\
5203560104  &  553  &  2022-11-20T09:25:34  &  2022-11-20T23:28:32  &  1824  &  2427728  \\
5203560105  &  750  &  2022-11-23T07:59:22  &  2022-11-23T08:12:27  &  986  &  37590  \\
5203560106  &  1460  &  2022-11-25T08:29:42  &  2022-11-25T12:47:47  &  6831  &  743269  \\
5203560107  &  2383  &  2022-11-29T15:38:20  &  2022-11-29T20:25:12  &  5568  &  826158  \\
5203560108  &  3685  &  2022-11-30T07:12:45  &  2022-11-30T23:20:02  &  8435  &  2785696  \\
5203560109  &  1597  &  2022-12-01T00:08:19  &  2022-12-01T23:28:11  &  3177  &  4031574  \\
5203560110  &  3042  &  2022-12-02T00:54:37  &  2022-12-02T22:47:23  &  4558  &  3780690  \\
5203560111  &  5369  &  2022-12-03T00:15:20  &  2022-12-03T23:32:31  &  15700  &  4023780  \\
5203560112  &  6626  &  2022-12-04T04:06:26  &  2022-12-04T22:54:30  &  25175  &  3248798  \\
5203560113  &  7727  &  2022-12-05T04:57:39  &  2022-12-05T21:58:49  &  19194  &  2940934  \\
5203560114  &  4854  &  2022-12-06T04:12:48  &  2022-12-06T16:41:45  &  7242  &  2156945  \\
5203560115  &  1677  &  2022-12-08T01:18:50  &  2022-12-08T23:05:59  &  4921  &  3764562  \\
5203560116  &  3000  &  2022-12-11T16:04:43  &  2022-12-11T22:30:43  &  4954  &  1111588  \\
5203560117  &  2720  &  2022-12-11T23:55:30  &  2022-12-12T12:26:39  &  4390  &  2163227  \\
5203560118  &  331  &  2022-12-13T13:07:14  &  2022-12-13T13:12:47  &  406  &  15950  \\
5203560120  &  1238 &  2023-01-04T03:06:01  &  2023-01-04T09:23:19  &  10006  &  1086573  \\
5203560121  &  4403 &  2023-01-05T03:51:23  &  2023-01-05T22:37:39  &  23652  &  3243567  \\
5203560122  &  3291  &  2023-01-05T23:58:24  &  2023-01-06T18:44:32  &  21511  &  3243276  \\
5203560123  &  381  &  2023-01-07T08:28:40  &  2023-01-07T08:35:01  &  3673  &  18270  \\
5203560124  &  1402  &  2023-01-07T23:56:24  &  2023-01-08T07:47:12  &  6344  &  1355881  \\
5203560125  &  1065  &  2023-01-09T03:47:23  &  2023-01-09T16:20:41  &  13851  &  2169533  \\
5203560126  &  835  &  2023-01-26T13:14:37  &  2023-01-26T14:53:32  &  1518  &  284835  \\
5203560127  &  1584  &  2023-01-27T01:40:27  &  2023-01-27T09:34:52  &  2465  &  1366209  \\
5203560128  &  904  &  2023-01-28T21:01:29  &  2023-01-28T22:41:52  &  1363  &  289074  \\
5203560129  &  1932  &  2023-01-30T10:11:11  &  2023-01-30T18:04:54  &  2575  &  1364212  \\
5203560130  &  1579  &  2023-01-31T04:45:04  &  2023-01-31T23:37:06  &  3085  &  3260241  \\
5203560131  &  617  &  2023-02-01T05:31:33  &  2023-02-01T05:41:50  &  929  &  29611  \\
5203560132  &  1025  &  2023-02-02T01:54:59  &  2023-02-02T17:20:38  &  2157  &  2665839  \\
5203560133  &  1994  &  2023-02-03T10:08:17  &  2023-02-03T19:40:06  &  3636  &  1646766  \\
5203560134  &  5750  &  2023-02-03T23:54:49  &  2023-02-04T23:36:01  &  11891  &  4092995  \\
5203560135  &  2356  &  2023-02-05T00:42:27  &  2023-02-05T22:49:03  &  4179  &  3820496  \\
5203560136  &  9150  &  2023-02-05T23:58:14  &  2023-02-06T23:50:08  &  42975  &  4123825  \\
5203560137  &  7017  &  2023-02-07T01:06:14  &  2023-02-07T23:02:12  &  33607  &  3789960  \\
5203560138  &  9054  &  2023-02-08T00:18:34  &  2023-02-08T23:49:45  &  245981  &  4064241  \\
\hline
\end{tabular}
}
\end{minipage}%
\begin{minipage}[t]{0.49\textwidth}
\centering
\resizebox{\textwidth}{!}{
\begin{tabular}{lrccrr}
\hline
\hline
ObsID & $T_{\rm exp}$ & \multicolumn{1}{c}{Start time} & \multicolumn{1}{c}{Stop time} & $N_{\rm cts}$ & $N_{\rm p}$ \\ 
      & (s)           & (UTC)                         & (UTC)                        &               &             \\
\hline
5203560139  &  4082  &  2023-02-09T01:12:05  &  2023-02-09T18:04:02  &  33591  &  2914359  \\
5203560140  &  1811  &  2023-02-10T06:06:54  &  2023-02-10T14:09:15  &  1537  &  34788  \\
5203560141  &  1479  &  2023-02-14T06:08:47  &  2023-02-14T06:36:48  &  2952  &  80648  \\
5203560142  &  2558  &  2023-02-16T10:45:03  &  2023-02-16T14:04:34  &  11291  &  574591  \\
5203560143  &  1080  &  2023-02-18T15:20:26  &  2023-02-18T17:09:08  &  2354  &  312975  \\
5203560144  &  786  &  2023-02-26T12:40:21  &  2023-02-26T12:53:27  &  2600  &  37719  \\
6203560101  &  101  &  2023-03-09T21:25:52  &  2023-03-09T21:28:07  &  75  &  6429  \\
6203560102  &  2227  &  2023-03-10T03:13:50  &  2023-03-10T23:29:27  &  7872  &  3500962  \\
6203560103  &  525  &  2023-03-11T02:26:41  &  2023-03-11T02:35:38  &  1539  &  25743  \\
6203560104  &  1275  &  2023-03-12T12:34:27  &  2023-03-12T17:22:43  &  3831  &  830162  \\
6203560105  &  1671  &  2023-03-15T20:45:08  &  2023-03-15T21:21:47  &  4273  &  105438  \\
6203560106  &  1688  &  2023-03-17T05:37:47  &  2023-03-17T15:12:50  &  3458  &  1656128  \\
6203560107  &  1139  &  2023-03-23T00:54:11  &  2023-03-23T01:13:10  &  2415  &  54593  \\
6203560109  &  1013  &  2023-04-04T12:46:35  &  2023-04-04T13:03:28  &  3437  &  48573  \\
6203560110  &  972  &  2023-04-06T12:47:56  &  2023-04-06T13:04:08  &  3076  &  46644  \\
6203560111  &  663  &  2023-04-08T14:23:05  &  2023-04-08T14:45:49  &  2769  &  65367  \\
6203560112  &  2529  &  2023-04-10T06:23:54  &  2023-04-10T11:18:09  &  7786  &  847410  \\
6203560114  &  6147  &  2023-04-12T00:10:34  &  2023-04-12T23:28:34  &  9878  &  4026061  \\
6203560115  &  730  &  2023-04-13T02:47:17  &  2023-04-13T18:13:42  &  1694  &  2668078  \\
6203560116  &  2323  &  2023-04-14T04:53:02  &  2023-04-14T18:53:34  &  3663  &  2420712  \\
6203560117  &  1989  &  2023-04-15T05:42:17  &  2023-04-15T10:31:54  &  5545  &  834051  \\
6203560118  &  867  &  2023-04-28T12:19:58  &  2023-04-28T14:14:25  &  5036  &  329566  \\
6203560119  &  1042  &  2023-05-12T19:57:58  &  2023-05-12T20:23:24  &  2049  &  73227  \\
6203560120  &  2888  &  2023-05-16T01:50:38  &  2023-05-16T23:39:48  &  6674  &  3770259  \\
6203560121  &  4547  &  2023-05-17T06:52:44  &  2023-05-17T23:52:57  &  8621  &  2938205  \\
6203560122  &  2925  &  2023-05-18T01:23:44  &  2023-05-18T11:15:49  &  7291  &  1705016  \\
6203560123  &  1026  &  2023-05-20T18:32:08  &  2023-05-20T18:59:43  &  5029  &  79421  \\
6203560124  &  1913  &  2023-05-22T15:29:32  &  2023-05-22T22:03:56  &  11604  &  65757  \\
6203560125  &  1329  &  2023-05-26T10:58:52  &  2023-05-26T12:43:03  &  804068  &  32256  \\
6203560136  &  225  &  2023-08-09T19:30:55  &  2023-08-09T19:35:45  &  539  &  13919  \\
6203560137  &  951  &  2023-08-12T07:29:13  &  2023-08-12T07:45:06  &  2486  &  45708  \\
6203560138  &  554  &  2023-08-16T18:25:39  &  2023-08-16T20:06:33  &  13578  &  290582  \\
6203560139  &  118  &  2023-10-17T14:49:28  &  2023-10-17T14:51:28  &  253  &  5748  \\
6203560141  &  1616  &  2023-10-21T10:04:33  &  2023-10-21T11:51:44  &  1754  &  35758  \\
6203560142  &  916  &  2023-10-23T10:04:28  &  2023-10-23T10:19:44  &  795  &  43918  \\
6203560143  &  653  &  2023-10-27T07:03:10  &  2023-10-27T07:14:03  &  1457  &  31336  \\
8203560101  &  783  &  2025-04-17T05:55:35  &  2025-04-17T07:34:31  &  8387  &  284899  \\
\hline
\end{tabular}
}
\end{minipage}

\medskip
\emph{Notes:} $N_{\rm cts}$: number of counts; $N_{\rm p}$: number of trial periods.
\end{table*}

\end{document}